\documentclass[fleqn,twoside]{article}
\usepackage{espcrc2}

\usepackage{graphicx}
\usepackage[figuresright]{rotating}

\newcommand{\AmS}{{\protect\the\textfont2
  A\kern-.1667em\lower.5ex\hbox{M}\kern-.125emS}}

\newcommand{\be}{\begin{equation}}
\newcommand{\ee}{\end{equation}}
\newcommand{\bea}{\begin{eqnarray}}
\newcommand{\eea}{\end{eqnarray}}
\newcommand{\bi}{\begin{itemize}}
\newcommand{\ei}{\end{itemize}}
\newcommand{\ben}{\begin{enumerate}}
\newcommand{\een}{\end{enumerate}}
\newcommand{\bt}{\begin{tabbing}}
\newcommand{\et}{\end{tabbing}}

\title{
   \vspace*{-80pt}
   {\normalsize \hfill {\sf KEK-CP-144}} \\
   {\normalsize \hfill {\sf UT-CCP-P140}} \\
   {\normalsize \hfill {\sf UTHEP-474}} \\
   \vspace*{20pt}
   Light hadron spectrum in three-flavor QCD
   with 
   $O(a)$-improved Wilson quark action
   \thanks{Talk presented by T.~Kaneko}
}

\author{
   CP-PACS and JLQCD Collaborations: 
   T.~Kaneko\address[KEK]{High Energy Accelerator Research Organization
                          (KEK), Tsukuba, Ibaraki 305-0801, Japan},
   S.~Aoki\address[Tsukuba]{Institute of Physics, University of 
                            Tsukuba, Tsukuba, Ibaraki 305-8571, Japan},
   M.~Fukugita\address[ICRR]{Institute for Cosmic Ray Research, 
                             University of Tokyo, 
                             Kashiwa 277-8582, Japan},
   S.~Hashimoto\addressmark[KEK]
   K-I.~Ishikawa\address[RCCP]{Center for Computational Physics, 
                               University of Tsukuba, Tsukuba, 
                               Ibaraki 305-8577, Japan},
   T.~Ishikawa\addressmark[RCCP],
   N.~Ishizuka$^{\rm b,d}$, 
   Y.~Iwasaki$^{\rm b,d}$, 
   K.~Kanaya$^{\rm b,d}$, 
   Y.~Kuramashi\addressmark[KEK], 
   M.~Okawa\address[Hiroshima]{Department of Physics, Hiroshima 
                               University, Higashi-Hiroshima, Hiroshima
                               739-8526, Japan}
   N.~Taniguchi\addressmark[Tsukuba], 
   N.~Tsutsui\addressmark[KEK], 
   A.~Ukawa$^{\rm b,d}$, 
   T.~Yoshi\'e$^{\rm b,d}$ 
}

\begin{document}

\begin{abstract}
   We report on a calculation of the light hadron spectrum 
   and quark masses in three-flavor dynamical QCD 
   using the non-perturbatively $O(a)$-improved Wilson quark action 
   and a renormalization-group improved gauge action. Simulations 
   are carried out on a $16^3 \times 32$ lattice at $\beta=1.9$, 
   where $a^{-1} \simeq 2$~GeV, 
   with 6 $ud$ quark masses corresponding 
   to $m_{\pi}/m_{\rho} \simeq 0.64$\,--\,0.77
   and 2 $s$ quark masses close to the physical value.
   We observe that the inclusion of dynamical strange quark brings 
   the lattice QCD meson spectrum to good agreement 
   with experiment. Dynamical strange quarks also
   lead to a reduction of the $uds$ quark masses by
   about 15\%. 
   
\end{abstract}

\maketitle

\section{Introduction}

As a joint collaboration of CP-PACS and JLQCD, we have initiated 
a project for the precise measurement of the light hadron spectrum 
and the quark mass with three-flavor full QCD. 
The systematic deviation of the quenched lattice spectrum 
from experiment found in \cite{Spectrum.Nf0.CP-PACS} 
is significantly reduced with the inclusion of dynamical up and down
quarks \cite {Spectrum.Nf2.CP-PACS,Spectrum.Nf2.JLQCD}. 
We now extend the work to three flavors.

\setcounter{footnote}{0}
We adopt the renormalization-group improved gauge action and 
the $O(a)$-improved Wilson quark action with
the non-perturbatively improved 
clover coefficient $c_{\rm SW}$  
obtained in \cite{NPcsw}.
Simulations are carried out at $\beta\!=\!1.9$, 
at the scale $a^{-1}\!=\!2.05(5)$~GeV,
on a $16^3\times32$ lattice
using a Polynomial Hybrid Monte Carlo algorithm
developed in \cite{PHMC}.
Six values of the $ud$ quark mass, corresponding to 
$m_{\rm PS,LL}/m_{\rm V,LL}\!\simeq\!0.64$\,--\,0.77\footnote{
   In this report, subscripts LL, LS and SS represent 
   the light-light, light-strange and strange-strange mesons,
   respectively.
},
are used to extrapolate to the physical value
$m_{\pi}/m_{\rho}\!\simeq\!0.18$.
Two values are taken for the $s$ quark mass, corresponding 
to $m_{\rm PS,SS}/m_{\rm V,SS}\!\simeq\!0.72$ and 0.77.
They are close to the semi-empirical value $m_{\eta_s}/m_{\phi}\!=\!0.68$ 
from chiral perturbation theory, so that a short extrapolation
is enough to get to the physical value.
3000 HMC trajectories are accumulated at each simulation parameter.

\section{meson masses}

We focus on the meson spectrum 
since our lattice size of $La\!\simeq\!1.6$~fm 
is too small for baryons \cite{Spectrum.Nf2.JLQCD}.


\begin{figure}[t]
   \begin{center}
      \includegraphics[width=0.84\linewidth,clip]{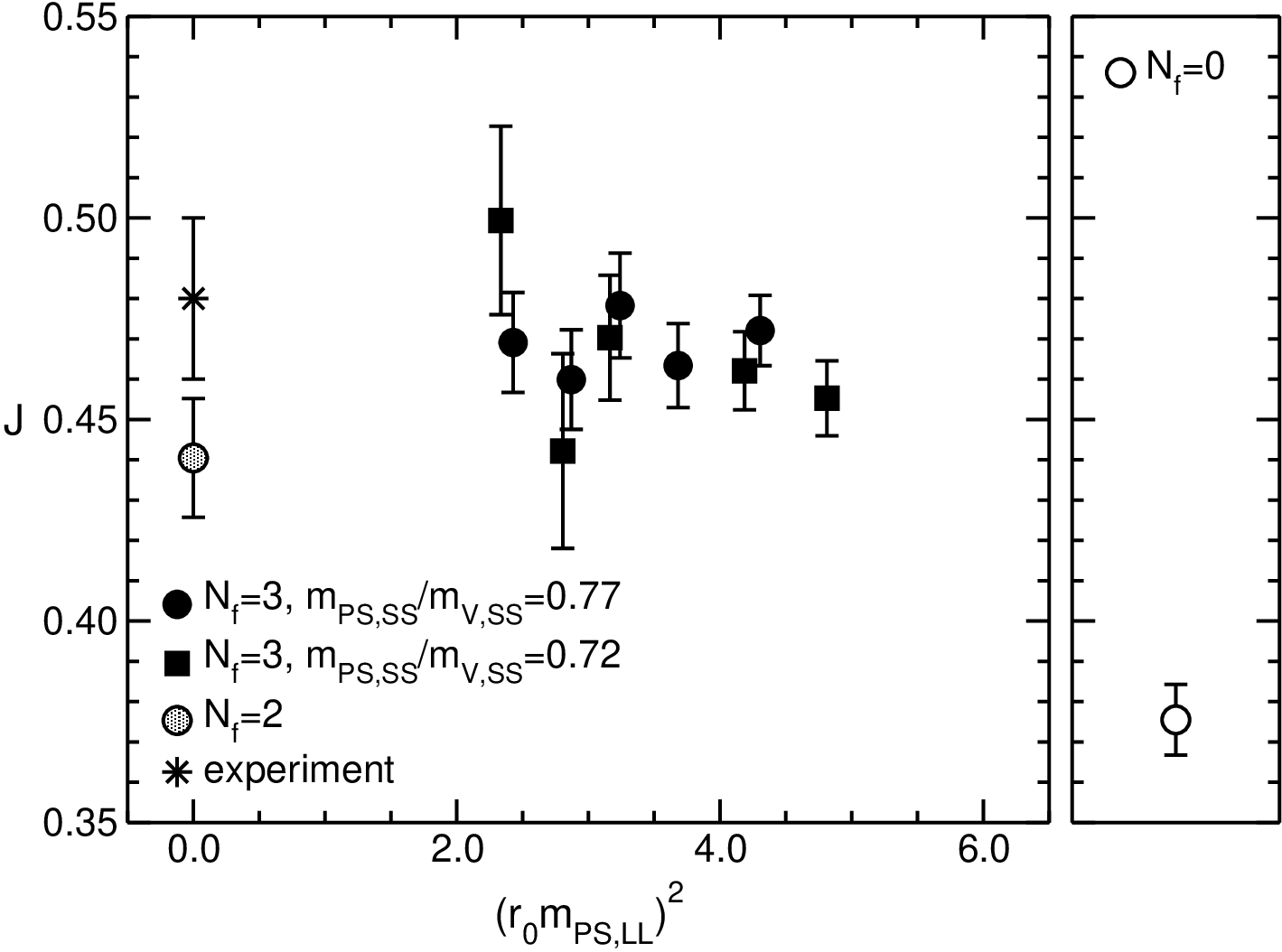}
   \end{center}
   \vspace*{-14mm}
   \caption{
      The $J$ parameter in two- and three-flavor QCD
      (left panel) and in quenched QCD (right panel).
   }
   \label{fig:had:J}
   \vspace*{-6mm}
\end{figure}

Fig.~\ref{fig:had:J} shows  
the $J$ parameter \cite{J} defined by 
\bea 
   J
 = m_{\rm V,LS}
   \frac{ m_{V,LS} -m_{V,LL}}
        { m_{PS,LS}^2 
         -m_{PS,LL}^2 }.
  \label{eqn:had:J_ratio}
\eea
This is compared with the previous results in two-flavor and quenched QCD
\cite{Spectrum.Nf2.CP-PACS}, which are evaluated 
for the physical kaons 
$m_{\rm PS}/m_{\rm V}\!=\!m_{K}/{m_{K^*}}$
as $ J \!=\! m_{\rm V} (d m_{\rm V}/ d m_{\rm PS}^2 )$
after taking the chiral limit for sea quarks\footnote{
We now prefer the definition of 
$J$ by Eq.~(\ref{eqn:had:J_ratio}),
as it is more straightforward, using the hadron masses measured
in the simulation without resorting to
any extrapolation procedures.}.
We see that $J$ in three-flavor QCD stays at values systematically higher than 
in quenched and two-flavor QCD, 
and is consistent with experiment.

Our lattice size $La\!\simeq\!1.6$~fm is significantly 
smaller than that used in quenched or two-flavor QCD.
Because finite-size effects generally lower the 
$J$ parameter \cite{Spectrum.Nf2.JLQCD}, we conclude 
that the consistency of 
$J$ between three-flavor QCD and experiment is not 
an artifact of finite-size effects.


Chiral extrapolations are carried out 
using polynomial forms.
%
We fix the lattice spacing and the $ud$ quark mass 
with $m_{\pi}$ and $m_{\rho}$.
The strange quark mass is fixed with either
$m_K$ ($K$-input) or $m_{\phi}$ ($\phi$-input).

\begin{figure}[t]
   \begin{center}
      \includegraphics[width=0.86\linewidth,clip]{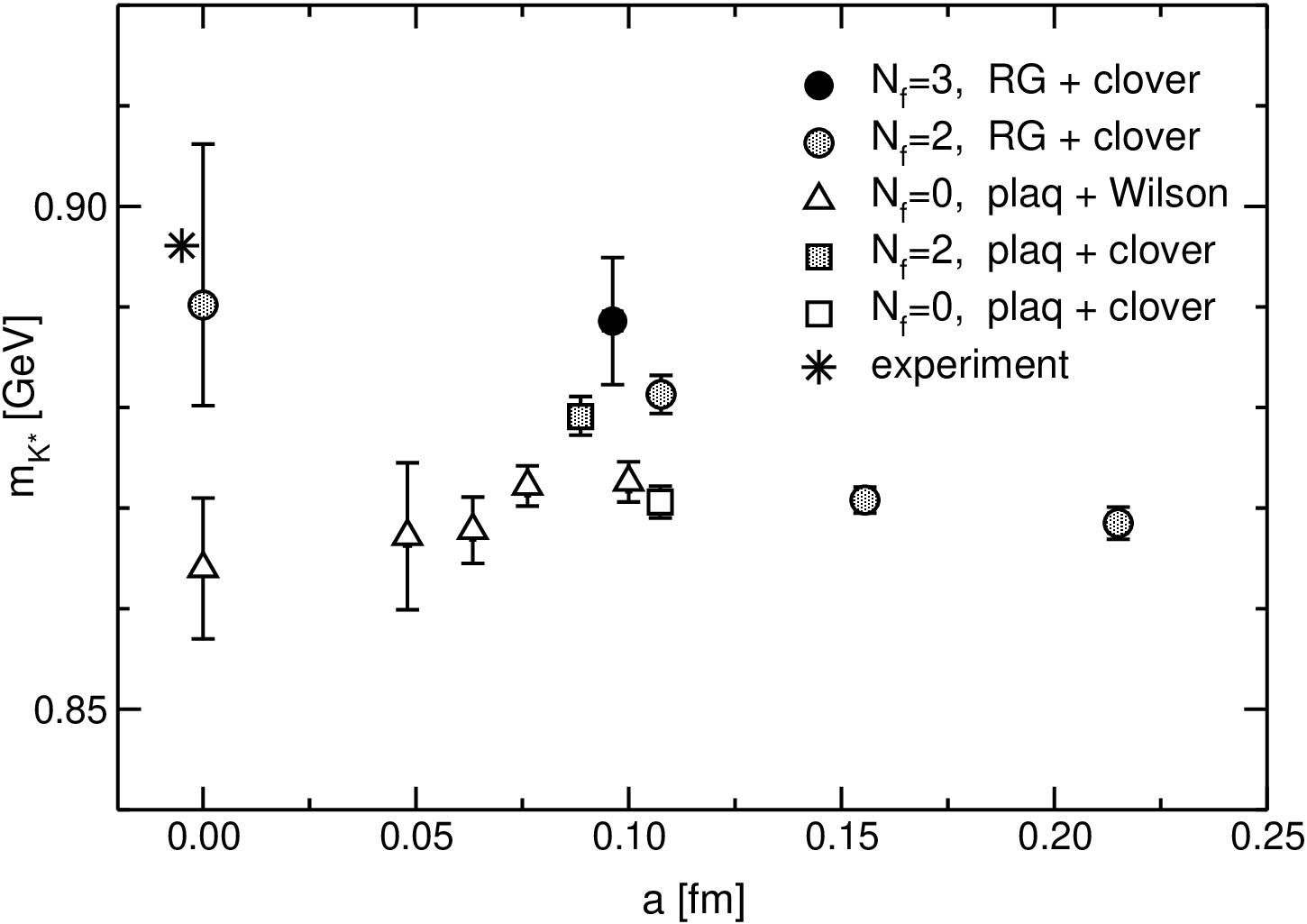}
   \end{center}
   \vspace{-14mm}
   \caption{
     $K^*$ meson mass as a function of lattice spacing.
   }
   \label{fig:had:scaling}
   \vspace*{-6mm}
\end{figure}

In Fig.~\ref{fig:had:scaling}, $m_{K^*}$ 
is plotted as a function of the lattice spacing.
The three-flavor result is higher than that in two-flavor QCD 
at comparable lattice spacings.
The consistency of the three-flavor value with experiment
seen already at a finite lattice spacing lends us hope
that scaling violation with the use of the non-perturbatively 
$O(a)$-improved action is indeed small and the consistency 
with experiment holds in the continuum limit.
This should be verified in a future study.

\begin{figure}[t]
   \begin{center}
      \includegraphics[width=0.82\linewidth,clip]{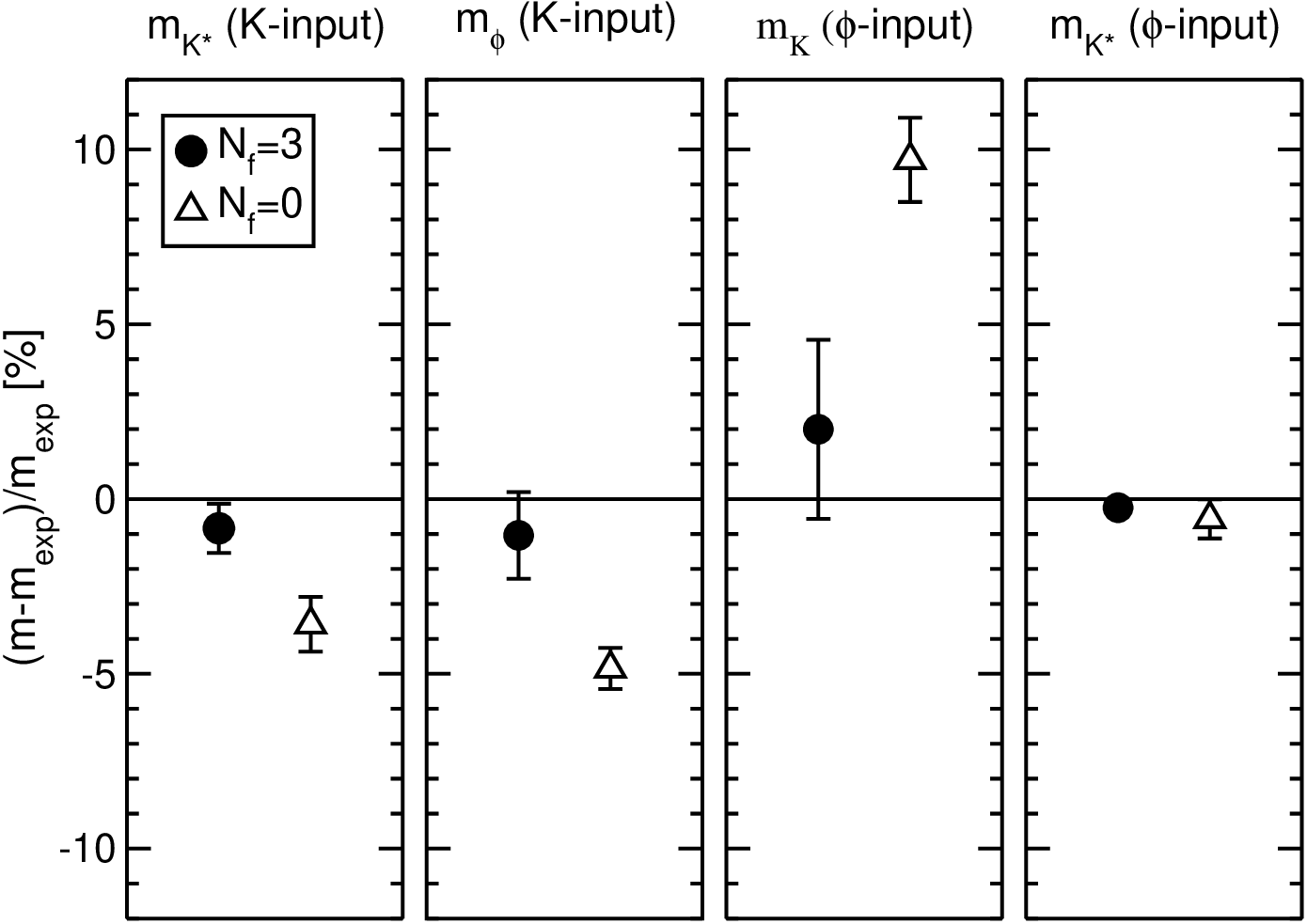}
   \end{center}
   \vspace*{-11mm}
   \caption{
     Deviation of meson spectrum from experiment 
     in three-flavor (filled symbols) and quenched QCD
     (open symbols).
   }
   \label{fig:had:meson}
   \vspace*{-5mm}
\end{figure}

Similar consistency with experiment can also be seen 
in other meson masses as shown in Fig.~\ref{fig:had:meson},
where deviations from experiment are plotted. 
Improvement over the quenched results of~\cite{Spectrum.Nf0.CP-PACS}
is evident, as anticipated from the agreement of the $J$
with experiment.

%
%

\section{Quark masses}

%

The physical quark masses,
$m_{ud}$ and $m_s$, are 
calculated through the axial-vector Ward identity (AWI):  
$m_q \!=\! \langle \Delta_4 A_4 P \rangle 
          /(2 \left\langle P P\right\rangle)$.
The matching to the $\overline{\rm MS}$ scheme 
is carried out at $\mu\!=\!a^{-1}$ 
using the one-loop $Z$ factor \cite{matching}
with the tadpole improvement.
The renormalized mass is evolved to $\mu\!=\!2$~GeV 
using the 4-loop $\beta$ function. 

The vector Ward identity (VWI) may offer 
an independent determination of $m_{ud}$ and $m_s$.
The critical hopping parameter $K_c$ for 
$ud$ quarks, defined where the light-light PS meson mass vanishes,
depends on $m_s$ due to the lack of chiral symmetry.
The two natural choices,  $K_c(m_s\!=\!0)$,
i.e., $K_c$ in degenerate three-flavor QCD, 
and $K_c(m_s=m_s^{\rm phys})$ with the physical $m_s$,
however, lead to $m_{ud}$ with 
the sign opposite to each other, indicating a
large uncertainty from
chiral symmetry breaking
at $a^{-1}\!\simeq\!2$~GeV.
The determination of the VWI $m_{ud}$ requires 
a careful study of scaling violation;
we focus on the AWI quark mass in this report.

\begin{figure}[t]
   \begin{center}
      \includegraphics[width=0.84\linewidth,clip]{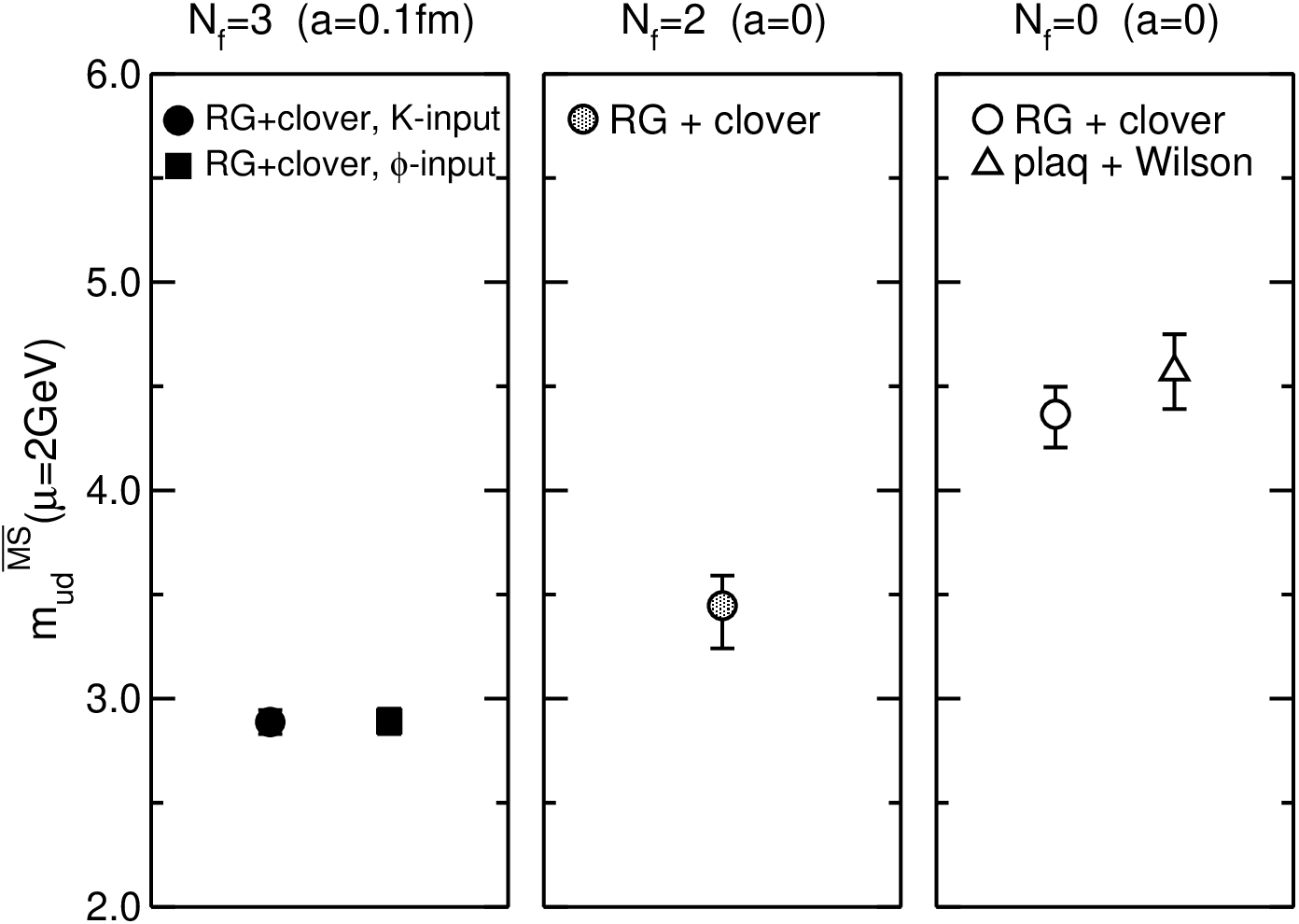}
   \end{center}
   \vspace*{-3mm}
   \begin{center}
      \includegraphics[width=0.84\linewidth,clip]{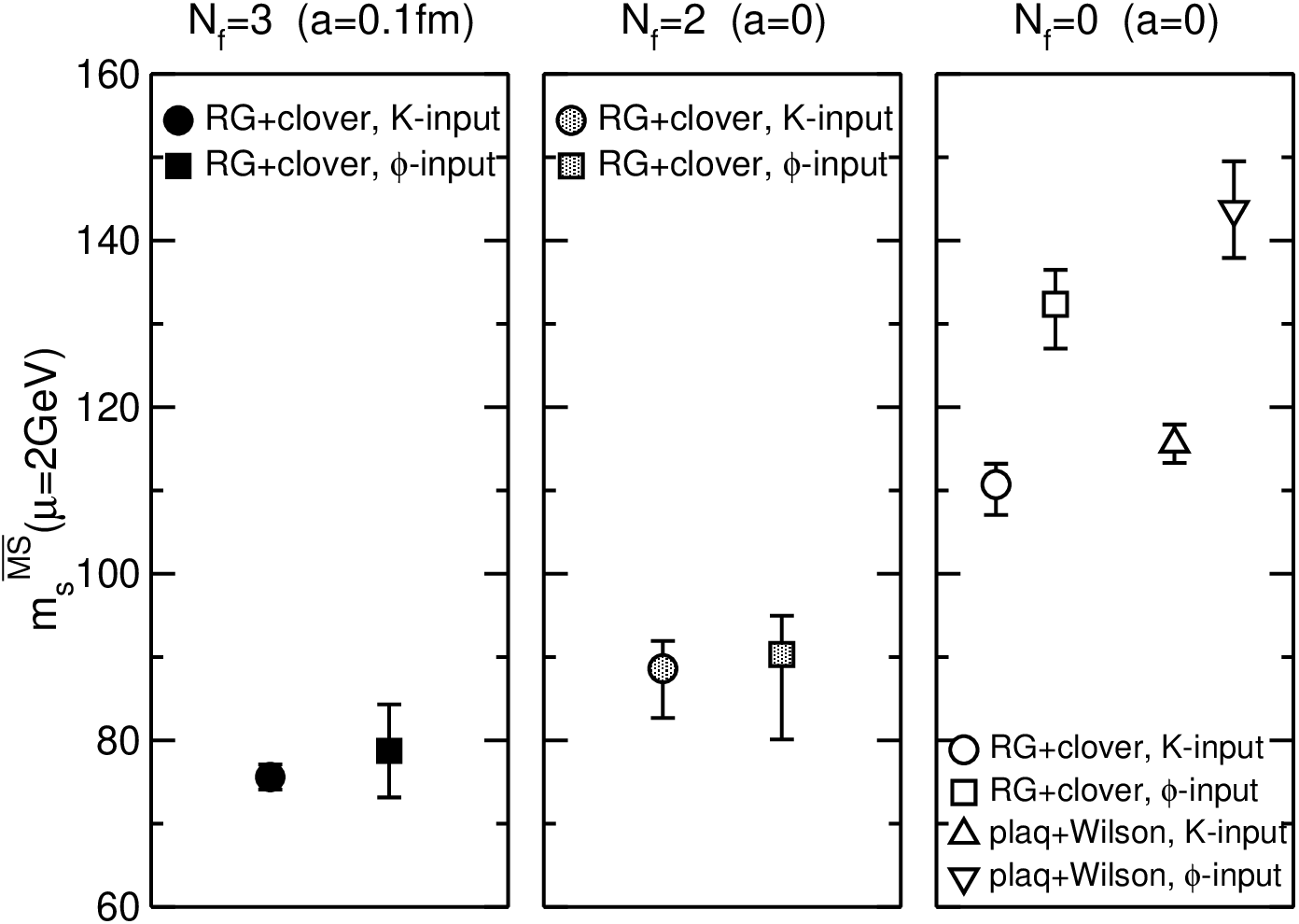}
   \end{center}
   \vspace*{-10mm}
   \caption{
     Comparison of $m_{ud}$ (top figure) and $m_s$ (bottom figure)
     between full and quenched QCD.
     The left, middle and right panels in each figure show results 
     in three- and two-flavor
     and quenched QCD, respectively.
   }
   \label{fig:mq:SQE}
   \vspace*{-3mm}
\end{figure}

Fig.~\ref{fig:mq:SQE} compares  $m_{ud}$ and $m_s$ in three-flavor QCD
with our previous results in 
two-flavor and quenched QCD
\cite{Spectrum.Nf0.CP-PACS,Spectrum.Nf2.CP-PACS}.
The discrepancy in $m_s$ between $K$- and $\phi$-inputs
in quenched QCD disappears in two- and three-flavor QCD.
Note that $m_{ud}$ and $m_s$
decrease with an increasing number of flavors of dynamical quarks. 

Taking the AWI mass with the $K$-input as the central value 
and including the deviation between the $K$- and $\phi$-inputs
into the error,
we obtain
\bea
   m_{ud}^{\overline{\rm MS}} 
   & = &
   2.89(6)~\mbox{MeV},
   \\
   m_{s}^{\overline{\rm MS}} 
   & = &
   75.6(3.4)~\mbox{MeV},
   \\
   m_{s}/m_{ud} 
   & = &
   26.2(1.0)
\eea
in three-flavor QCD. Dynamical strange quarks decrease
both $m_{ud}$ and $m_s$ by about 15\%.
The ratio $m_{s}/m_{ud}$ is 
consistent with the one-loop estimate of 
chiral perturbation theory 24.4(1.5) \cite{mq.ChPT}.

%

\section{Conclusion}

Our simulation for three-flavor QCD shows that
the inclusion of dynamical strange quarks brings the
lattice QCD meson spectrum to good agreement with experiment.
A non-negligible reduction is also found for quark masses.
These preliminary results are sufficiently encouraging to warrant
further detailed and careful investigations with time-consuming
three-flavor full QCD.

\vspace{3mm}

This work is supported by Large Scale Simulation Program No. 98
(FY2003) of High Energy Accelerator Research Organization (KEK),
and also in part by the Grants-in-Aid of the Ministry of Education
(Nos. 12740133, 13135204, 13640259, 13640260, 14046202, 14540289, 
14740173, 15204015, 15540251, 15540279).
NT is supported by JSPS.

\end{document}